\documentclass[aps,pre,twocolumn,twoside, amsmath,amssymb,
 aps,bbm,floatfix,showpacs]{revtex4-1}

\usepackage{xcolor}
\usepackage{graphicx}
\usepackage{dcolumn}
\usepackage{bm}
\usepackage{framed}
\usepackage{hyperref}
\usepackage{dsfont}
\usepackage[normalem]{ulem}
\usepackage{nicefrac}

\begin{document}

\title{Assessing the quantumness of a damped two-level system}
\author{Alexander Friedenberger}
 
\author{Eric Lutz}%

\affiliation{Department of Physics, Friedrich-Alexander-Universit\"at Erlangen-N\"urnberg, D-91058 Erlangen, Germany}

\begin{abstract}
We perform a detailed analysis of the nonclassical properties of a damped two-level system. We  compute and compare three different criteria of quantumness, the $l_1$-norm of coherence, the Leggett-Garg inequality and a quantum witness based on the no-signaling in time condition. We show that all three quantum indicators decay exponentially in time as a result of the coupling to the  thermal reservoir. We further demonstrate that the corresponding characteristic times are identical and given by the coherence half-life. These results quantify how violations of Leggett-Garg inequalities and nonzero values of the quantum witness are connected to the coherence of the  two-level system.

\end{abstract}

\maketitle

\section{Introduction} The question of what genuinely distinguishes quantum from classical physics is as old as quantum theory itself \cite{Einstein1935,Schroedinger1935}. That issue is not only of fundamental, but also of technological importance. Quantum features have indeed been shown to be a  physical resource that allows  to perform tasks that are not possible classically \cite{nie00}. Prominent examples include quantum cryptography \cite{Ekert1991}, quantum teleportation \cite{Bennett1993} and quantum computing \cite{sho94}. However, the boundary between the classical and the quantum world is notoriously blurry \cite{zur91,zur03}. Characterizing  nonclassicality is, as a result, a challenging exercise. The topic has recently attracted renewed interest in the fields of quantum biology \cite{lam13,moh14}, quantum computation \cite{zag14,ron14} and quantum thermodynamics \cite{Uzdin2015,fri15}.

An inherent property of quantum mechanics is the coherent superposition of different states \cite{wal08}. A popular example of such a quantum superposition is given by  Schr\"odinger's famous cat  that can  simultaneously be both dead and alive \cite{Schroedinger1935}. Cat states are nowadays routinely created and studied in the lab \cite{bru96,mya00}. A rigorous theoretical framework of quantum coherence as a physical resource has been developed lately \cite{Aberg2006, Plenio2014,yao15,str15,Winter2016} (see Ref.~\cite{str16} for a  review). In particular, different quantifiers of coherence, such as coherence measures and monotones, have been introduced \cite{str16}. A  commonly used example of a coherence monotone is the $l_1$-norm, $C_{l_1}(\rho) = \sum_{i\neq j} |\rho_{ij}|$, which is simply the sum of the modulus of the nondiagonal matrix elements of the density operator $\rho$ of a system \cite{Plenio2014}. A nonvanishing $C_{l_1}(\rho) $ indicates the presence of quantum coherence in the considered basis.

 Another approach to identify quantumness is to impose classical constraints that are violated by quantum theory. For instance, the classical assumptions of realism and locality lead to Bell's inequality \cite{Bell1964}. After early successful observations of the violation of the Bell inequality by quantum systems \cite{Clauser1974,Aspect1981,Aspect1982}, loophole-free  experiments have been reported recently \cite{Hanson2015,Zeilinger2015,Shalm2015}. The violation of Bell's inequality is  related to the existence of nonclassical spatial correlations between two parties. It is therefore not well--suited to detect quantum behavior of a single system. The Leggett-Garg inequality, on the other hand, may be seen as a temporal analogue of Bell's inequality  \cite{Leggett1985,Emary2014}. It is based  on the classical notions of macroscopic realism, i.e. the assumption that macroscopic systems remain in a well--defined state at all times, and of noninvasive measurability, i.e. the possibility to measure in principle the state of a system without perturbing it. A violation of the Leggett-Garg inequality reveals the presence of nonclassical temporal correlations in the dynamics of an individual  system. Such a violation has been experimentally seen in an increasing number of systems in the last  years, from superconducting qubits to neutrinos \cite{Korotkov2010,Goggin2011,Xu2011,Dressel2011,Suzuki2012,Waldherr2011,yon12,George2013,Athalye2011,Souza2011,Katiyar2013,Knee2012,Zhou2015,rob15,for16}. It should, however, be pointed out that while the nonviolation of Bell's inequalities is  necessary and sufficient for local realism \cite{Fine1982}, the nonviolation of  Leggett-Garg inequalities is a necessary but not sufficient condition for macroscopic realism \cite{Kofler2016}.

 A third witness for nonclassical behavior has been  proposed latterly, based on the classical assumption of no-signaling in time, i.e.  the idea that a measurement does not change
the outcome statistics of a later measurement \cite{Nori2012,Kofler2012} (it has also been called non-disturbing-measurement condition \cite{George2013,mar14}).
 In essence, the criterion compares the dynamics of the population of a quantum system in the presence and in the absence of a measurement. A deviation between the two dynamics then signifies quantumness. Advantages of the quantum witness are that i) its implementation only requires two time measurements, in contrast to the three measurements usually needed to test the Leggett-Garg inequality and that ii) it   involves one-point expectations,  rather
than two-point correlations. In addition, it provides a necessary and sufficient condition for macrorealism \cite{Kofler2016}. Experimental implementations with a single atom \cite{rob15}  and  a superconducting flux qubit have been reported \cite{kne16}.

In this article, we compare the above three criteria of nonclassicality by applying them to a damped two-level system, a paradigmatic model of quantum optics \cite{wal08} and condensed-matter physics \cite{wei08}. We begin by solving the Markovian master equation of the dissipative two-level system in the Heisenberg picture and by computing the two-time correlation function of the Pauli operator $\sigma_x$ in Section II. We then evaluate the coherence monotone $C_{l_1}(\rho)$ in the $\sigma_z$-basis in Section III, derive the Leggett-Garg inequalities in Section IV and calculate the  quantum witness in Section IV. We find that all three quantum indicators show  an exponential decay of the quantum properties of the two-level system induced by the coupling to the external thermal reservoir. Remarkably, we establish that all three characteristic times are equal and given by the coherence half-life  \cite{cor12,pau13}. These findings clarify how violations of the Leggett-Garg inequalities and non-vanishing values of the quantum witness are   related to the quantum coherence of the two-level system.

\section{Damped two-level system} 
We consider a two-level system weakly coupled to a bath of harmonic oscillators at temperature $T$, e.g. a two-level atom interacting with a thermal radiation field. In the Born-Markov approximation, the time evolution in the Heisenberg picture of a system observable $X(t)$ is governed by a Lindblad master equation  of the form \cite{Breuer,Alicki} (we set $\hbar=1$ throughout for simplicity),
\begin{equation}
\begin{aligned}
\frac{dX}{dt}&=\mathcal{L}[X]=i\frac{\omega}{2} \left[\sigma_z,X\right]+\frac{\partial X}{\partial t} \\
&+{\frac{\gamma_{0}}{2}n(\omega,T)}\left(\sigma_{-}\left[X,\sigma_{+}\right] +\left[\sigma_{-},X\right]\sigma_{+}\right)\\
&+{\frac{\gamma_{0}}{2}(n(\omega,T)+1)}\left(\sigma_{+}\left[X,\sigma_{-}\right] +\left[\sigma_{+},X\right]\sigma_{-}\right).\label{eq:mastereq}
\end{aligned}
\end{equation}
Here $\omega$ is the frequency of the two-level system, $\gamma_0$  the spontaneous decay rate and $n(\omega,T)= [\exp(\omega/k_B T)-1]^{-1}$ the thermal occupation number. We denote the total transition rate by $\gamma=\gamma_0\left[2 n(\omega,T)+1\right]$.

The master equation \eqref{eq:mastereq} may be conveniently solved by using a superoperator formalism \cite{Crawford1958,muk95}.  By choosing a basis consisting of the Pauli matrices $\sigma_i\, (i=x ,y, z)$ and the identity operator $I$,  the matrix representation of the Liouvillian superoperator  $\mathcal{L}$ reads,
\begin{equation}
\frac{d}{dt}\begin{pmatrix}
 \sigma_x\\
\sigma_y\\
\sigma_z\\
I
\end{pmatrix}= \mathcal{L}\begin{pmatrix}
\sigma_x\\
\sigma_y\\
\sigma_z\\
I
\end{pmatrix}= 
{\begin{pmatrix}
-\frac{\gamma}{2}&-\omega&0&0\\
\omega&-\frac{\gamma}{2}&0&0\\
0&0&-\gamma&-\gamma_0\\
0&0&0&0
\end{pmatrix}}\begin{pmatrix}
\sigma_x\\
\sigma_y\\
\sigma_z\\
I
\end{pmatrix}.\label{eq:Mastereq}
\end{equation}
This solution may be used to compute two-point correlation functions of the system with the help of the quantum regression theorem \cite{Carmichael}. For an operator $O(t)$, we have, 
\begin{equation*}
\frac{d}{d\tau}\langle O(t)\sigma_j(t+\tau)\rangle=\sum_i \mathcal{L}_{ji}\langle O(t)\sigma_i(t+\tau)\rangle.
\end{equation*}
For the   time-symmetrized  correlation function $C(\tau)=\langle\{\sigma_x(t),\sigma_x(t+\tau)\}\rangle/2$, we find (see Appendix A), \begin{equation}
C(\tau)=\exp\left(-\frac{\gamma}{2}\tau\right)\cos\left(\omega\tau\right).\label{eq:Correlation}
\end{equation}
The  correlation function $C(\tau)$ only depends on the  time lag $\tau$ and is hence stationary. It will be useful in the derivation of the Leggett-Garg inequalities in Sect. IV. 

\section{$l_1$-norm of coherence}
In this Section, we assess the quantumness of the two-level system by computing the $l_1$-norm of coherence $C_{l_1}=\sum_{i\neq j}|\rho_{ij}|$, arguably the simplest and most intuitive coherence monotone \cite{Plenio2014}. This quantity depends  on the state of the system, described by the density operator $\rho$,  and on the basis in which the matrix elements $\rho_{ij}$ are evaluated. We here choose the natural basis for the problem at hand given by the eigenbasis of the Hamiltonian of the two-level system $H= (\omega/2) \sigma_z$. In the $\sigma_z$ basis, the density matrix $\rho$ may be expressed in terms of the expectation values of the Pauli operators  \cite{Breuer}:

\begin{equation}
\rho=\frac{1}{2}\begin{pmatrix}I+\langle\sigma_z\rangle &\langle \sigma_x\rangle -i \langle \sigma_y \rangle\\
 \langle \sigma_x\rangle +i \langle \sigma_y \rangle&I-\langle\sigma_z\rangle. \label{eq:density}
 \end{pmatrix}.
\end{equation}
We accordingly obtain the coherence monotone, 
\begin{equation}
C_{l_1}=\sqrt{\langle\sigma_x\rangle^2+\langle \sigma_y\rangle^2}.\label{eq:l1}
\end{equation}
Equation \eqref{eq:l1} may be computed from the solution \eqref{eq:Mastereq} of the master equation \eqref{eq:mastereq}.
We select the initial state of the two-level system to be maximally coherent. As shown in Ref.~\cite{Plenio2014}, an example of such a state is given by,
\begin{equation}
|\Psi_d\rangle=\frac{1}{\sqrt{d}}\sum_{i=1}^{d}|i\rangle,
\end{equation}
where $d$ is the dimension of the system. We choose the eigenstate $|+\rangle =\left(|\uparrow\rangle +|\downarrow\rangle\right) /\sqrt{2} $ of the Pauli operator $\sigma_x$, which is obviously of this form for $d=2$. As a result, we obtain the  $l_1$-norm,
\begin{equation}
C_{l_1}=\exp({-{\gamma \tau}/{2}}). \label{eq:coherencemonotone}
\end{equation}
In general, the  $l_1$-norm of coherence  satisfies the inequality $C_{l_1}\le d-1$ \cite{str16}. The coherence monotone \eqref{eq:coherencemonotone}  takes its maximum possible value at $\tau=0$ from which it decays exponentially with a characteristic timescale given by the coherence time $t_c = 2/\gamma$. The latter quantity is defined as the time at which  coherence is reduced to $1/e$ times its initial value \cite{Breuer}. It is also advantageous to introduce the coherence half-life, $\tau_c = 2\ln 2/\gamma$, defined as the time at which coherence decays to half its initial value \cite{cor12,pau13}.

\section{Leggett-Garg inequality}
 Let us next characterize the nonclassicality of the two-level system by using the Leggett-Garg inequality. Consider a  dichotomous observable $Q$, which can take the values $\pm1$, and which  is measured at three consecutive times $t_1$, $t_2$ and $t_3$. Based on the classical assumptions of macroscopic realism and noninvasive measurability, one can  derive the  inequality \cite{Leggett1985,Emary2014},
\begin{equation}
K_3=\langle Q(t_2) Q(t_1)\rangle+\langle Q(t_3)Q(t_2) \rangle-\langle Q(t_3)Q(t_1)\rangle \le 1.\label{eq:LGI}
\end{equation}
Quantum theory violates the above  inequality. A value of the Leggett-Garg function $K_3$ above unity is thus a signature of nonclassical behavior. For  $Q= \sigma_x$ and equally spaced measurements with separation $\tau/2$, we find,
\begin{equation}
K_3(\tau)=-C(\tau)+2C(\tau/2), \label {9}
\end{equation}
where $C(\tau)$ is the quantum correlation function \eqref{eq:Correlation}. Equation \eqref{9} is shown in Fig.~1 for the case of an isolated  two-level system ($\gamma=0$). We observe that the function $K_3$ takes on values that are larger than one (shaded area), revealing quantum properties. However, owing to the oscillatory nature of $K_3$, we also note values that are smaller than one, even  though the dynamics of the system is coherent in the absence of the thermal reservoir. The dynamics of the  system may thus be quantum even though the Leggett-Garg inequality is not violated.

The above-mentioned problem can here  be solved  by considering Leggett-Garg-type inequalities introduced in Ref.~\cite{Huelga1995}. Based on the classical assumption of macroscopic realism and the condition of stationarity, the following inequalities hold for Markovian systems:
\begin{eqnarray}
K_{+}(\tau)&=-C(\tau)-2C(\tau/2)\le 1 \label{eq:LGIstat1}\\
K_{-}(\tau)&=-C(\tau)+2C(\tau/2)\le 1.\label{eq:LGIstat2}
\end{eqnarray}
These inequalities are violated by unitary quantum dynamics; note that $K_-(\tau)=K_3(\tau)$ \cite{rem}.  As seen in Fig.~1, the two inequalities for $K_+$ and $K_-$ are complementary: one being violated when the other one is not, and vice versa.  The addition of a second Leggett-Garg function therefore allows a complete detection of the nonclassical  properties of the two-level system for $\gamma=0$. An experimental violation of the two inequalities \eqref{eq:LGIstat1} and \eqref{eq:LGIstat2} has been described  in Refs.~\cite{Waldherr2011,yon12,Zhou2015}.

\begin{figure}
\includegraphics[width=0.46\textwidth]{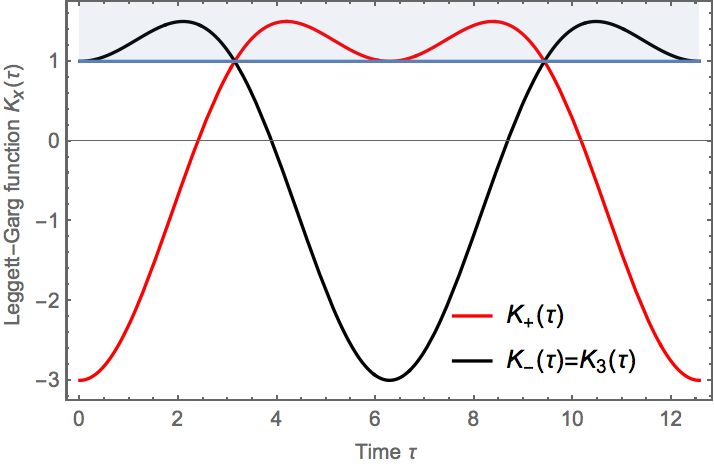}
\caption{(color online) Leggett-Garg functions   $K_+(\tau)$ (red) and $K_3(\tau) = K_-(\tau)$ (black), Eqs.~(9)-(11),  as a function of the time  $\tau$ for an isolated two-level system ($\gamma=0$). The blue shaded area corresponds to the classically forbidden regime indicated by a violation of the Leggett-Garg inequalities (9)-(11). The frequency is set to $\omega=1$ as in  the other figures.}
\label{fig:LGI}
\end{figure}
\begin{figure}
\includegraphics[width=0.46\textwidth]{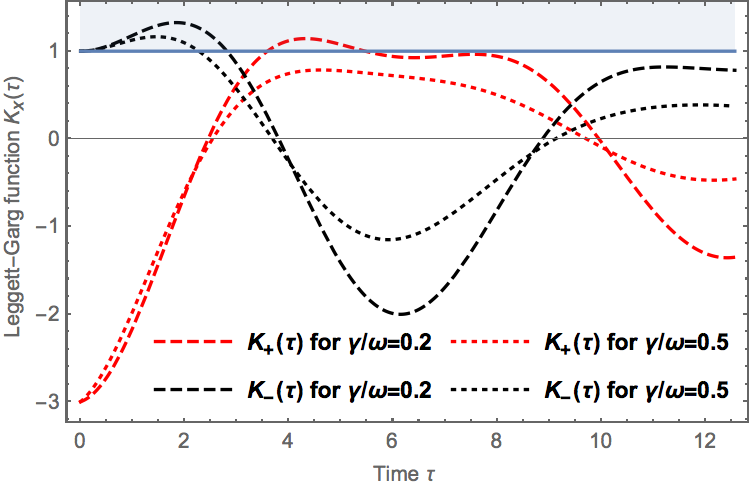}
\caption{(color online) Leggett-Garg functions $K_+(\tau)$ and $K_-(\tau)$, Eqs.~(10)-(11),  as a function of the time  $\tau$ for a damped two-level system for two values of the damping coefficient $\gamma$. Above a critical  value of the time $\tau$, violations of the Leggett-Garg inequalities (10)-(11) are not possible and the dynamics will be classical.}
\end{figure}
\begin{figure}[b]
\includegraphics[width=0.49\textwidth]{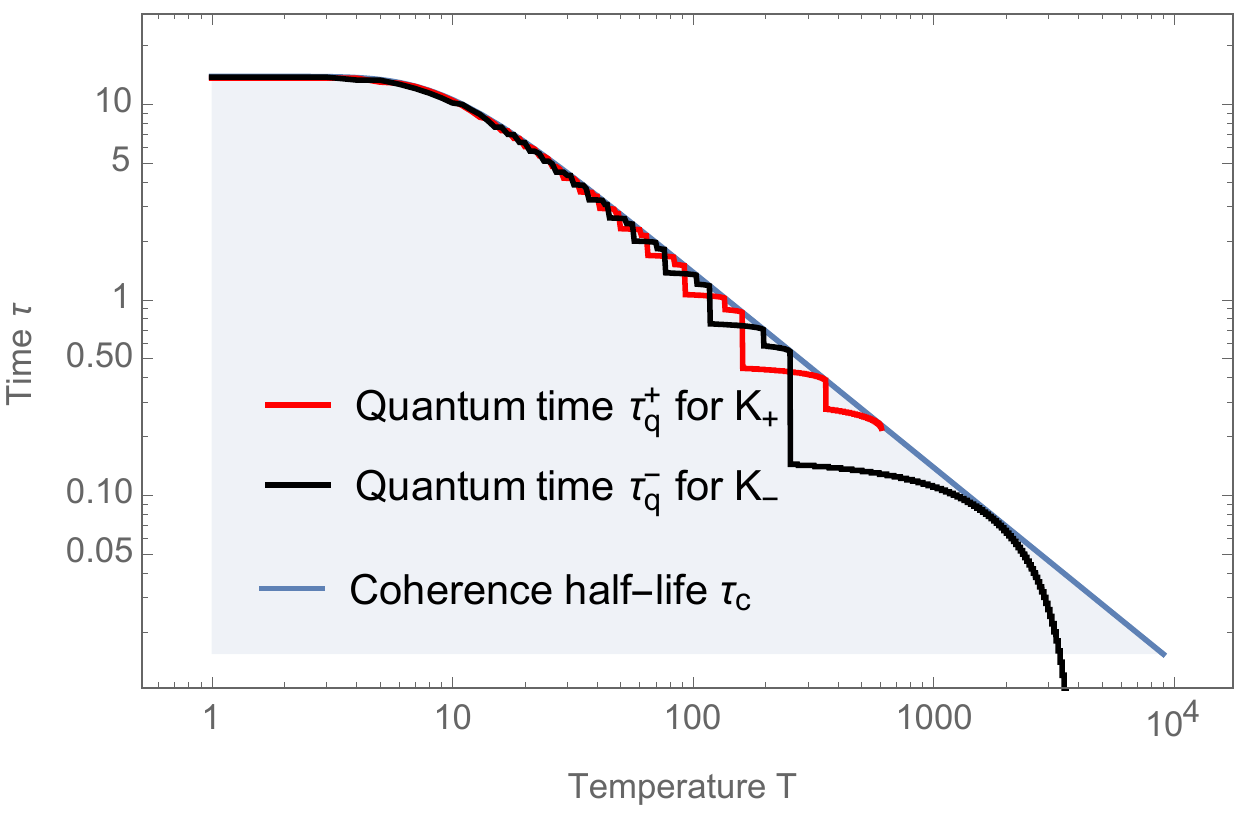}
\caption{(color online) Quantum times $\tau_q^\pm$ (red and black), Eq.~(12), defined as the maximum time $\tau$ for which a violation of the Leggett-Garg inequalities (10)-(11) is possible as a function of the temperature $T$ of the thermal reservoir. Both quantum times $\tau_q^\pm$ are bounded from above by the decoherence half-life $\tau_c$ (blue), Eq.~(7).  Parameters are here $\omega=20$ and $\gamma_0=0.01$.}
\end{figure}
Figure 2 displays the Leggett-Garg functions $K_+(\tau)$ and $K_-(\tau)$ for increasing values of the coupling constant $\gamma$. The interaction with the thermal reservoir leads to a damping of the oscillations of the two functions. After a certain maximal measurement spacing, no further violations of the Leggett-Garg inequalities \eqref{eq:LGIstat1}-\eqref{eq:LGIstat2} will be observed and the dynamics of the two-level system will be classical. We accordingly introduce a quantum time $\tau_q^\pm$, defined as the largest measurement time $\tau$, between the first and the last measurement, for which a violation of the inequalities \eqref{eq:LGIstat1}-\eqref{eq:LGIstat2} may occur:
\begin{equation}
\tau_q^\pm=\text{max}  \left\{ \tau | K_\pm(\tau)\ge 1\right\}.
\end{equation}
The times $\tau_q^\pm$ characterize the quantum-to-classical transition of the two-level system, as quantum features are only possible for times smaller than $\tau_q^\pm$.

Figure 3 shows the numerically determined quantum times $\tau_q^\pm$  as a function of the temperature $T$ of the thermal reservoir.
As expected the quantum times decrease with increasing temperature, indicating that nonclassical properties (shaded area) are destroyed faster when the system is coupled to a hot environment. We remark that the times $\tau_q^\pm$ are step functions (owing to the oscillatory nature of the Leggett-Garg functions $K_+$ and $K_-$) and that the decay with temperature of the height of the steps is precisely given by the coherence half-life $\tau_c = 2\ln 2/\gamma$ of the coherence monotone $C_{l_1}$, Eq.~\eqref{eq:coherencemonotone}.  We may hence conclude that violations of the Leggett-Garg inequalities (10)-(11)   occur until the coherence induced by  a first measurement has decayed to half its initial value at the time of a subsequent measurement.

\section{Quantum witness} 
\begin{figure}[t]
\includegraphics[width=0.49\textwidth]{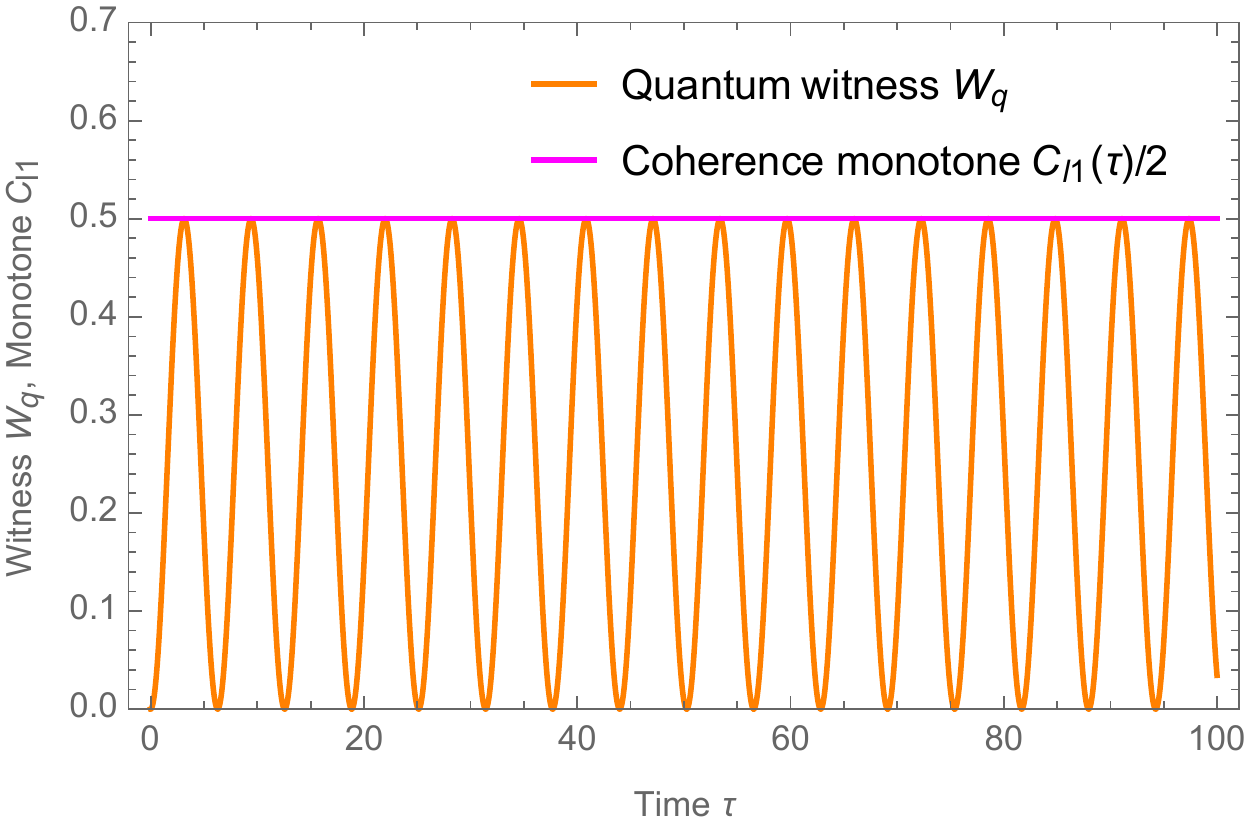}
\caption{(color online) Quantum witness $\mathcal{W}_q$ (orange), Eq.~(14),  as a function of the time  $\tau$ for an isolated two-level system ($\gamma=0$). The quantum witness $\mathcal{W}_q$   is bounded from above by the coherence monotone $C_{l_1}(\tau)/2$ (purple), Eq.~(7).}
\end{figure}

We finally analyze the quantum properties of the damped two-level system by employing the quantum witness which has been introduced in two slightly different ways in Refs.~\cite{Nori2012,Kofler2012}.
 Following Ref.~\cite{Nori2012}, we consider a  $d$-level system and denote by  $p_n(t_0)$   its probability to be at $t=t_0$ in the classical  state $n$ ($ 1\leq n\leq d$). The probability to find the system in state $m$ at time $t$ is   \cite{Nori2012},
 \begin{equation}
\bar p_m (t)=\sum_{n=1}^d \Omega_{mn}(t,t_0)p_n(t_0) \label{eq:ProbEvo},
\end{equation}
 where the propagator $\Omega_{mn}(t,t_0)=p(m,t|n,t_0)$ gives the probability of a transition from state $n$ to state $m$ in time $t-t_0$ (the bar emphasizes that the state $n$ is  classical). The quantum witness is  then defined as \cite{Nori2012},
 \begin{equation}
\mathcal{W}_q=\left| p_m(t)-\sum_{n=1}^d \Omega_{mn}(t,t_0)p_n(t_0)\right |. \label{14}
\end{equation}
 A nonzero value of the quantum  witness, $\mathcal{W}_q>0$, reveals the nonclassicality of the initial state. Compared to the Leggett-Garg inequality \eqref{eq:LGI}, the condition of noninvasive measurability is here replaced by the requirement to perform ideal state preparation of each state $n$ and $m$.

 \begin{figure}[t]
\includegraphics[width=0.49\textwidth]{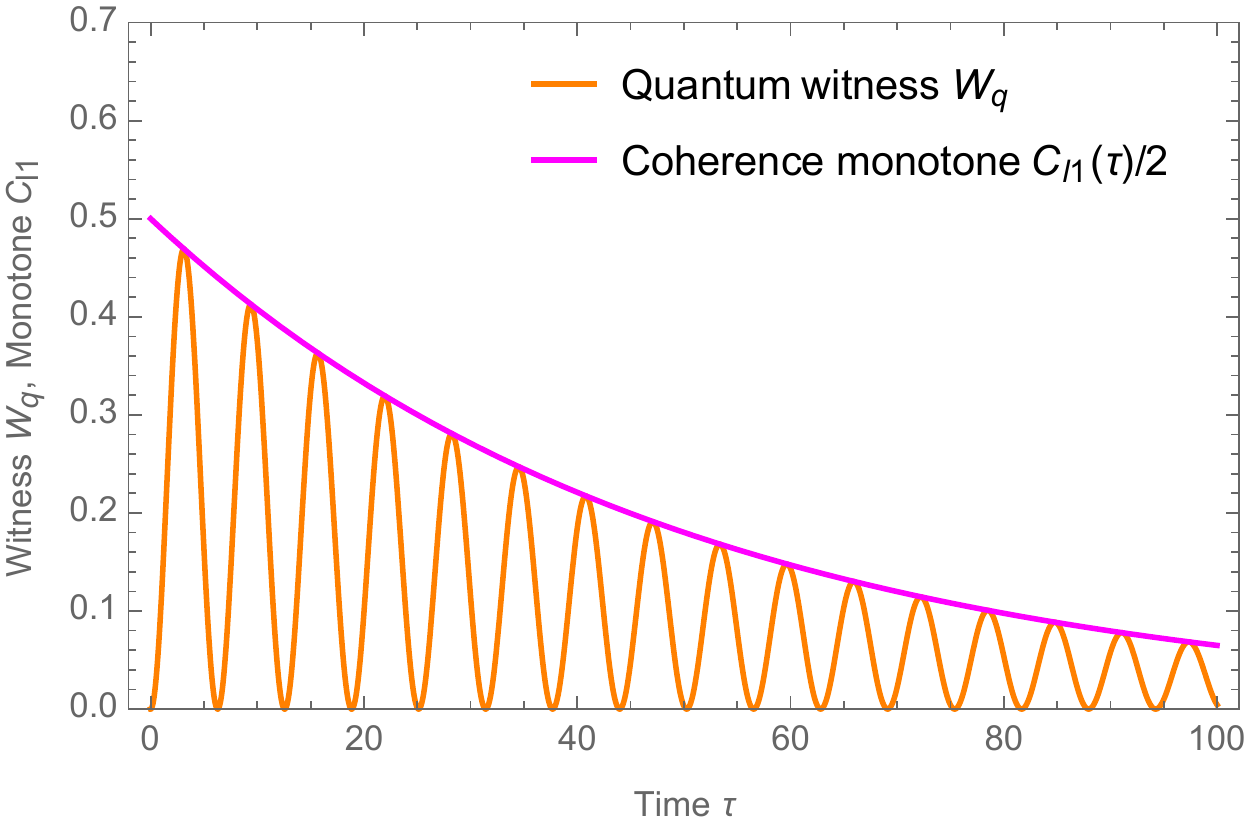}
\caption{(color online) Quantum witness $\mathcal{W}_q$ (orange), Eq.~(14),  as a function of the time  $\tau$ for a damped two-level system ($\gamma=0.04$). The quantum witness $\mathcal{W}_q$   is bounded from above by the  coherence monotone $C_{l_1}(\tau)/2$ (purple), Eq.~(7).}
\label{fig:Witness}
\end{figure}

The same expression may be directly obtained from the classical no-signaling in time condition \cite{Kofler2012}. Consider two observables $A$ and $B$, respectively measured at time $t=t_0$ and at time $t>0$. The measurement outcome $n$ of $A$ is obtained with probability $p_n(t_0)$, while the measurement outcome $m$ of $B$ is obtained with probability $p'_m(t)$. For a joint measurement of the two observables, the probability of obtaining $m$ in the second measurement is given by,
\begin{equation}
p'_m(t)=\sum_{n=1}^d p(m,t |n,t_0)p_n(t_0),\label{eq:Prob}
\end{equation}
with the conditional probability $p(m,t |n,t_0)$.  In the absence of the first measurement on $A$, the probability of outcome $m$ of $B$ is denoted by $p_m(t)$. According to the classical no-signaling in time assumption, the measurement of  $A$ should have no influence on the statistical outcome of the later measurement of  $B$, and $p'_m(t) = p_m(t)$. The quantum witness is then defined as the difference $\mathcal{W}_q=| p_m(t)-p'_m(t)|$, which is identical to Eq.~\eqref{14}.

The quantum witness \eqref{14} may be evaluated for the dissipative two-level system by using the solution \eqref{eq:Mastereq} of the master equation \eqref{eq:mastereq}. The system is initially prepared in the $|+\rangle$ state at $t=0$. At time $t=\tau/2$, a first nonselective measurement in the $\sigma_x$-basis is performed or not, while at time $t=\tau$ the projector $\Pi_{+}=|+\rangle\langle +|$ is measured. By explicitly computing the propagator $\Omega_{mn}(\tau,\tau/2)$, we find (see Appendix B),
\begin{equation}
\mathcal{W}_q =\frac{1}{2}e^{-\gamma \tau/2}\sin^2(\omega \tau/2). \label{16}
\end{equation}
In the case of unitary dynamics ($\gamma= 0$), the quantum witness \eqref{16} reaches its maximum value $\mathcal{W}_q^\text{max} =1-{1}/{d} = 1/2$ at times  $\omega \tau=n\pi $ with$ \:n\in \mathbb{N}$ \cite{Emary2015}. We note that the maximum value $\mathcal{W}_q^\text{max}$ is equal to $C_{l_1}(\tau)/2$ (see Fig.~4). Moreover,  at times $\omega \tau=n\pi $, the two-level system is in a superposition of eigenstates of $\sigma_y$ and is therefore maximally disturbed by a measurement of $\sigma_x$ (see Appendix C). For nonunitary dynamics ($\gamma\neq0$), the maxima of the quantum witness decay exponentially in time with a characteristic time again equal  to the coherence half-life $\tau_c = 2 \ln2 /\gamma$ of the coherence monotone $C_{l_1}$, Eq.~\eqref{eq:coherencemonotone}. We further observe that $\mathcal{W}_q(\tau)\le C_{\ell_1}(\tau)/2$ and that the latter upper bound corresponds exactly  to the envelop of the oscillatory quantum witness (see Fig.~5).

We finally remark that the quantum witness is here directly related to the expectation value of the $\sigma_y$ operator,
\begin{equation}
\mathcal{W}_q(\tau)=\frac{1}{2}\langle \sigma_y \left(\tau/2\right)\rangle^2.
\end{equation}
This is an interesting result that may simplify the experimental detection of the nonclassicality of a damped two-level system: Instead of two measurements in the $\sigma_x$-basis required to realize the quantum  witness,  single measurements of  $\sigma_y$ after  a suitable state preparation along $\sigma_x$ should be sufficient.

\section{Summary}
We have presented a comprehensive examination of the quantum signatures of a damped two-level system. We have derived explicit expressions for the $l_{1}$-norm of coherence $C_{l_1}$, the Leggett-Garg functions $K_+$ and $K_-$ and the quantum witness $\mathcal{W}_q$ based on the no-signaling in time condition. We have shown that all three quantum indicators allow to identify a clear boundary between quantum and classical behavior, defined by a unique characteristic time given by the coherence half-life $\tau_c$.  This clarifies in a quantitative manner how violations of the Leggett-Garg inequalities and nonzero values of the quantum witness are  linked to the existence of coherence in the system. There exists, however, a crucial qualitative difference between the three quantifiers of nonclassicality. The coherence half-life characterizes the exponential temporal decay of both the $l_1$-norm of coherence and the quantum witness; it thus corresponds to a soft border between quantum and classical properties. By contrast, the coherence half-life defines a sharp transition  for the Leggett-Garg inequalities beyond which quantum features abruptly disappear, akin to so-called sudden-death  behaviors \cite{alm07}.  We finally mention that each of the three tests of quantumness faces different experimental challenges: full state tomography for the $l_{1}$-norm of coherence, the noninvasive measurement of two-time correlation functions for the Leggett-Garg inequalities, and ideal state preparation for the quantum witness. The noted direct connection (17) of  the quantum witness to the expectation value of the $\sigma_y$ Pauli operator may simplify such an experimental test.

\textit{Acknowledgments.} This work was partially supported
by the EU Collaborative Project TherMiQ (Grant Agreement
618074) and the COST Action MP1209.

\begin{appendix}

\section{Derivation of the correlation function }
In this section, we compute the two-point correlation function $C(\tau)$, Eq.~(3).
According to the quantum regression theorem,  the equation of motion for the two-time correlation function is the same as that for the corresponding one-time function in the limit of weak coupling \cite{Breuer,Carmichael}.  We thus have in general for operators $O$ and $A$,
\begin{equation}
\frac{d}{d\tau}\langle \hat{O}(t)\vec{A}(t+\tau)\rangle=\mathcal{L}\langle \hat{O}(t)\vec{A}(t+\tau)\rangle.
\end{equation}
Considering only the upper left submatrix of the superoperator $\mathcal{L}$, Eq.~(2), we find,
\begin{equation}
\label{eq}
\frac{d}{d\tau}\vec{C}=\begin{pmatrix}-\frac{\gamma}{2}&-\omega\\
\omega&-\frac{\gamma}{2}\end{pmatrix}\vec{C}(\tau),
\end{equation}
where the correlation vector $\vec{C}(\tau)$ is defined as, 
\begin{equation}
\vec{C}(\tau)=\begin{pmatrix}\langle \sigma_x(t)\sigma_x(t+\tau)\rangle\\\langle \sigma_x(t)\sigma_y(t+\tau)\rangle\end{pmatrix}.
\end{equation}
The solution to Eq.~\eqref{eq} is given by,
\begin{equation}
\vec{C}(\tau)=\left(
\begin{array}{cc}
 e^{-\frac{\gamma \tau}{2} } \cos ( \omega \tau ) & -e^{-\frac{\gamma t}{2} } \sin( \omega \tau ) \\
e^{-\frac{\gamma \tau}{2} } \sin ( \omega \tau )  & e^{-\frac{\gamma t}{2} } \cos ( \omega \tau )  \\
\end{array}
\right)\vec{C}(0),
\end{equation}
with the initial condition, 
\begin{equation}
\vec{C}(0)=\begin{pmatrix}\langle \sigma_x \sigma_x\rangle (t)\\ \langle \sigma_x \sigma_y\rangle (t)\end{pmatrix}=\begin{pmatrix} 1\\ i\langle \sigma_z\rangle (t)\end{pmatrix}.
\end{equation}
The last equality is a result of the  algebraic properties of the Pauli operators. The time-symmetrized correlation function $C(\tau)=\langle\{\sigma_x(t),\sigma_x(t+\tau)\}\rangle/2$ is equal to the real part of the above correlation function and reads,
\begin{equation}
C(\tau)=\exp\left({-\frac{\gamma}{2}\tau}\right )\cos\left(\omega \tau\right).
\end{equation}

\section{Calculation of the quantum witness}
In this section, we evaluate the quantum witness $\mathcal{W}_q$, Eq.~(16). In order to first compute the propagator $\Omega$, it is convenient to express the Liouville superoperator  $\mathcal{L}$ in a basis consisting of the projectors $\Pi_{\pm}$ onto  the $\sigma_x$ eigenstates $|\pm\rangle$ and the Pauli operators 
$\sigma_x$ and $\sigma_y$ (instead of the basis $\sigma_x,\sigma_y, \sigma_z$ and $I$ used previously). 
In that basis, the master equation (1) takes the  form,
\begin{equation}
\frac{d}{dt}\begin{pmatrix}\Pi_{+}\\ \Pi_{-}\\ \sigma_y \\ \sigma_z \end{pmatrix}= \mathcal{ L}\begin{pmatrix}\Pi_{+}\\ \Pi_{-}\\ \sigma_y \\ \sigma_z \end{pmatrix} = \begin{pmatrix}-\frac{\gamma}{4}&\frac{\gamma}{4}&-\frac{\omega}{2}&0\\ \frac{\gamma}{4}&-\frac{\gamma}{4}& \frac{\omega}{2}&0 \\ \omega&-\omega &-\frac{\gamma}{2}&0\\ -\gamma_0& -\gamma_0&0&-\gamma \end{pmatrix}\begin{pmatrix}\Pi_{+}\\ \Pi_{-}\\ \sigma_y \\ \sigma_z \end{pmatrix} \label{b1}
\end{equation}
The formal solution of Eq.~\eqref{b1} is,
\begin{widetext}
\begin{equation}
e^{\mathcal{L}\cdot t}=\left(
\begin{array}{cccc}
 \frac{1}{2} \left(1+e^{-\frac{ \gamma t}{2} }\cos ( \omega t)\right) & \frac{1}{2} \left(1-e^{-\frac{\gamma t}{2}} \cos ( \omega t)\right) & -\frac{1}{2} e^{-\frac{ \gamma t}{2} } \sin (\omega t) & 0 \\
 \frac{1}{2} \left(1-e^{-\frac{\gamma t}{2} } \cos ( \omega t)\right) & \frac{1}{2} \left(1+e^{-\frac{\gamma t}{2}} \cos ( \omega t)\right) & \frac{1}{2} e^{-\frac{\gamma t}{2} } \sin ( \omega t) & 0 \\
 e^{-\frac{\gamma t}{2}   } \sin (\omega t) & -e^{-\frac{\gamma t}{2}} \sin ( \omega t) & e^{-\frac{\gamma t}{2} } \cos ( \omega t ) & 0 \\
 \frac{\left(-1+e^{- \gamma t}\right) \text{$\gamma_0 $}}{\gamma } & \frac{\left(-1+e^{- \gamma t}\right) \text{$\gamma_0 $}}{\gamma } & 0 & e^{- \gamma t} \\
\end{array}
\right).\label{eq:Propagator}
\end{equation}
\end{widetext}
Assuming that the  system is initially at $t=0$ in  state $|+\rangle$, the time evolution of the $\Pi_+$ operator follows as,
\begin{equation}
\Pi_{+}(t)=\frac{1}{2} \left(1+e^{-\frac{\gamma  t}{2} } \cos ( \omega t)\right)\label{eq:Projector}.
\end{equation}
The (quantum) probability  $p_+(\tau)$ is then simply the expectation $\langle \Pi_+(\tau) \rangle$. On the other hand, the propagator $\Omega_{mn}(\tau,{\tau}/{2})$ with $(m,n)= (+,-)$ is described by the upper left $2\times 2$ matrix  of the full propagator,  Eq.~(\ref{eq:Propagator}):
\begin{equation}
\Omega\left(\tau,\frac{\tau}{2}\right)= \frac{1}{2}\begin{pmatrix} 1+e^{-\frac{ \gamma  \tau}{4} }\cos ( \omega \frac{\tau}{2}) &  1-e^{-\frac{ \gamma \tau}{4}} \cos ( \omega \frac{\tau}{2})  \\
 1-e^{-\frac{\gamma \tau}{4} } \cos ( \omega \frac{\tau}{2}) &  1+e^{-\frac{\gamma \tau}{4}} \cos ( \omega \frac{\tau}{2}) \end{pmatrix},
\end{equation}
The (classical) probabilities to find the two-level system in the states $|\pm\rangle$ is at $t={\tau}/{2}$ are further given by, 
\begin{equation}
\begin{pmatrix} p'_+\\ p'_-\end{pmatrix}\left(\frac{\tau}{2}\right)=\Omega\left(\frac{\tau}{2},0\right)\begin{pmatrix} p'_+\\ p'_-\end{pmatrix}(0).
\end{equation}
Combining these two expressions, the classical probability to find the system in state $|+\rangle$ at $t=\tau$ is,
\begin{equation}
p'_+(\tau)=\frac{1}{2} \left[1+e^{-\frac{\gamma  t}{2} } \cos ^2\left(\frac{ \omega t}{2}\right)\right]\label{eq:PropDist}.
\end{equation}
Equations \eqref{eq:Projector} and \eqref{eq:PropDist} finally lead to the witness,
\begin{equation}
\mathcal{W}_q=|\langle \Pi_+(\tau)\rangle -p'_+ (\tau)|=\frac{1}{2}e^{-\frac{\gamma}{2} \tau}\sin^2\left(\frac{\omega}{2} \tau \right).
\end{equation}

\section{Maximal measurement disturbance}
For an isolated two-level system $(\gamma=0)$, the quantum witness $\mathcal{W}_q$, Eq.~(16), reaches it maximal value when $\omega \tau=n\pi $ with$ \:n\in \mathbb{N}$ \cite{Emary2015}. This condition corresponds to a Larmor precession of the system from  the initial $\sigma_x$ eigenstate to (a mixture of) eigenstates of $\sigma_y$ prior to the first measurement. It is  intuitively clear that measuring $\sigma_x$ while the system is in a $\sigma_y$-eigenstate will lead to a strong disturbance. We will here show that the disturbance is maximal by evaluating the Hilbert-Schmidt norm of the commutator of these observables. We consider two  normed operators $A,B$ of a two-level system: 
\begin{equation}
\begin{aligned}
A=\frac{1}{\sqrt{2}}\left(\alpha_0 I +\sum_{i=1}^3 \alpha_i \sigma_i\right),
B=\frac{1}{\sqrt{2}}\left(\beta_0 I +\sum_{j=1}^3 \beta_j \sigma_j\right)
\end{aligned}
\end{equation}
The Hilbert-Schmidt norm $\left\lVert A \right\rVert_2^2$ is then simply \cite{nie00},
\begin{equation}
\left\lVert A \right\rVert_2^2= \text{Tr}(A^\dagger A) =\sum_{i=0}^3 |\alpha_i|^2.
\end{equation}
On the other hand, the commutator of $A,B$ reads
\begin{equation}
[A,B]=\sum_{i,j=1}^3\alpha_i \beta_j[\sigma_i,\sigma_j], \label{c3}
\end{equation}
with $[\sigma_i,\sigma_j]=\sum_k 2i\varepsilon_{ijk}\sigma_k$, where $\varepsilon_{ijk}$ is the Levi-Civita symbol. We have here used  the bilinearity of the commutator. By further introducing a new set of coefficients, $\lambda_k=2i\sum_{i,j=1}^3\alpha_i \beta_j \varepsilon_{ijk}$, we may write Eq.~\eqref{c3} as,
\begin{equation}
[A,B]=\sum_{k=1}^3\lambda_k\sigma_k.
\end{equation}
The Hilbert-Schmidt  norm of the commutator follows as,
\begin{equation}
\left\lVert [A,B] \right\rVert_2^2=\sum_{k=1}^3|\lambda_k|^2.
\end{equation}
This expression is proportional to the squared modulus of the cross product of $A,B$, if we interpret the operators as vectors with components $A_i=\alpha_i$, $B_i=\beta_i$. For fixed norms of the operators $A,B$, the maximum value is obtained by choosing orthogonal operators, corresponding to a vanishing  Hilbert-Schmidt product $\langle A,B \rangle =\text{Tr}(A^\dagger B)=0
$. For the normed operators $\sigma_x/\sqrt{2}$ and $\sigma_y/\sqrt{2}$, we find $\langle \sigma_x,\sigma_y\rangle/2=\text{Tr}(\sigma_x\sigma_y)/2=0$ and $\left\lVert \left[\sigma_x/\sqrt{2},\sigma_y/\sqrt{2}\right]\right\rVert_2^2=2$. The operators $\sigma_x$ and $\sigma_y$ are thus maximally incompatible.

 \begin{figure}[t]
\includegraphics[width=0.49\textwidth]{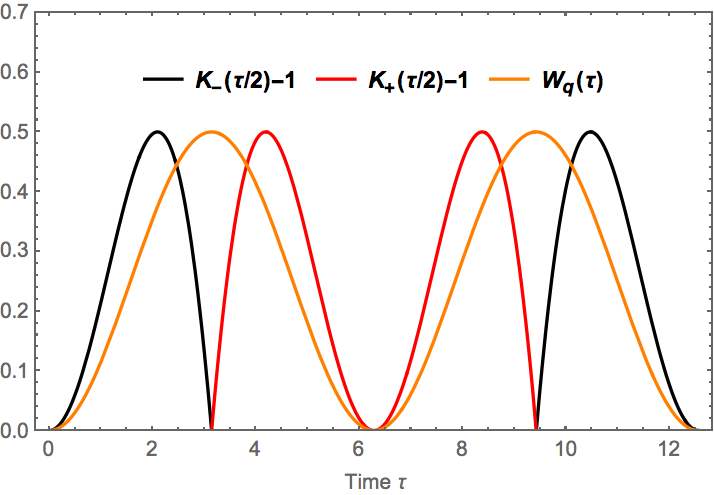}
\caption{(color online) Comparison between the quantum witness $\mathcal{W}_q$ (orange), Eq.~(14), and the Leggett-Garg functions $K_+$ (red) and $K_-$ (black), Eqs.~(10)-(11), for an isolated two-level system ($\gamma=0$). Maximal measurement disturbance coincides with maxima of the  witness and the quantum/classical boundary, $K_\pm=1$, of the Leggett-Garg functions. }
\label{fig:Witness}
\end{figure}

We additionally observe that the maxima of the witness correspond  to  Leggett-Garg functions $K_\pm(\tau)=1$, that is, to the  quantum/classical boundary (see Fig.~6). The Leggett-Garg inequality indeed quantifies measurement induced correlations, while the quantum witness quantifies measurement disturbance. Measurement disturbability is a prerequisite to induce correlations by a measurement. However, the maximally disturbing measurement, which corresponds to measuring $\sigma_x$ in a $\sigma_y$ eigenstate, does not lead to correlations between measurements. Any such measurement will yield either eigenvalue of $\sigma_x$ with equally probability, independent of the state preparation/initial measurement. This point thus corresponds to  the quantum/classical boundary of the Leggett-Garg functions.  We finally note that the Leggett-Garg inequality and the quantum witness here identify the same nonclassical domain. 

\end{appendix}

\end{document}